\documentclass[fleqn,usenatbib]{mnras}

\usepackage{newtxtext,newtxmath}

\usepackage[T1]{fontenc}

\DeclareRobustCommand{\VAN}[3]{#2}
\let\VANthebibliography\thebibliography
\def\thebibliography{\DeclareRobustCommand{\VAN}[3]{##3}\VANthebibliography}

\usepackage{graphicx}	
\usepackage{amsmath}	
 \usepackage{xcolor}

\def\heii{\ion{He}{II}}

\def\nii{\ion{N}{II}}

\def\oiii{\ion{O}{III}}

\def\sii{\ion{S}{II}}

\def\Ha{H$\alpha$}

\def\Gaia{{\it Gaia}}

\def\TESS{{\it TESS}}
\def\WISE{{\it WISE}}
\def\kms{{\>\rm km\>s^{-1}}}

\newcommand{\logg}{\mbox{$\log g$}}
\newcommand{\Teff}{\mbox{$T_\mathrm{eff}$}}

\title[Central Star of Fr 2-30]{Spectroscopic survey of faint planetary-nebula nuclei. II\null. The subdwarf~O central star of Fr 2-30}
\author[H. E. Bond et al.]{
Howard E. Bond,$^{1,2}$\thanks{E-mail: heb11@psu.edu}
Klaus Werner,$^{3}$ 
Gregory R. Zeimann,$^{4}$
and Jonathan Talbot$^{5}$
\\
$^{1}$Department of Astronomy \& Astrophysics, Pennsylvania State University, University Park, PA 16802, USA\\
$^{2}$Space Telescope Science Institute, 3700 San Martin Dr., Baltimore, MD 21218, USA\\
$^{3}$Institut f\"ur Astronomie und Astrophysik, Kepler Center for Astro and Particle Physics, Eberhard Karls Universit\"at, Sand 1, D-72076 T\"ubingen, Germany \\
$^{4}$Hobby-Eberly Telescope, University of Texas, Austin, Austin, TX 78712, USA \\
$^{5}$Stark Bayou Observatory, 1013 Conely Cir., Ocean Springs, MS 39564, USA
}

\date{Accepted XXX. Received YYY; in original form ZZZ}

\pubyear{2023}

\begin{document}
\label{firstpage}
\pagerange{\pageref{firstpage}--\pageref{lastpage}}
\maketitle

\begin{abstract}
Fr~2-30 = PN?\,G126.8$-$15.5 is a faint emission nebula, hosting a 14th-mag central star { that we identify here for the first time}. Deep \Ha\ and [\oiii] images reveal a roughly elliptical nebula with dimensions of at least $22'\times14'$, fading into a surrounding network of even fainter emission. Optical spectrograms of the central star show it to have a subdwarf~O spectral type, with a \Gaia\/ parallax distance of 890~pc. A model-atmosphere analysis gives parameters of $T_{\rm eff}=60,000$~K, $\log g= 6.0$, and a low helium content of $n_{\rm He}/n_{\rm H} = 0.0017$. 
The location of the central star in the $\log g$ -- $\log T_{\rm eff}$ plane is inconsistent with a post-asymptotic-giant-branch evolutionary status. Two alternatives are that it is a helium-burning post-extreme-horizontal-branch object, or a hydrogen-burning post-red-giant-branch star. In either case the evolutionary ages are so long that a detectable planetary nebula (PN) should not be present. 
We find evidence for a variable radial velocity (RV), suggesting that the star is a close binary. However, there are no photometric variations, and the spectral-energy distribution rules out a companion earlier than M2~V\null. 
The RVs of the star and surrounding nebula are discordant, and the nebula lacks typical PN morphology. We suggest that Fr~2-30 is a ``PN mimic''---the result of a chance encounter between the hot sdO star and an interstellar cloud. However, we note the puzzling fact that there are several nuclei of genuine PNe that are known to be in evolutionary states similar to that of the Fr~2-30 central star.
\end{abstract}




\section{Introduction} \label{sec:intro}

This is the second in a series of papers presenting results from a spectroscopic survey of central stars of faint planetary nebulae (PNe). It is being carried out with the second-generation Low-Resolution Spectrograph (LRS2; \citealt{Chonis2016}) of the 10-m Hobby-Eberly Telescope (HET; \citealt{Ramsey1998,Hill2021}), located at McDonald Observatory in west Texas, USA\null. An overview of the survey, a description of the instrumentation and data-reduction procedures, target selection, and some initial results were presented in our first paper \citep[][hereafter Paper~I]{Bond2023}. 

PNe are formed when a star ejects its outer layers and leaves the asymptotic-giant branch (AGB) in the HR diagram. When the remnant stellar core has evolved to a sufficiently high temperature, its ultraviolet (UV) radiation photoionises the surrounding ejecta, creating the optically visible PN\null. For recent reviews of PNe, see \citet{FrewParker2010} and \citet{Kwitter2022}. 

The targets of our LRS2 survey are central stars of nebulae with very low surface brightnesses. Numerous new faint PNe have been discovered in recent years, many of them by amateurs, as detailed in Paper~I\null. Our survey is aimed at characterising their faint central stars and pointing out those worthy of further investigation. In this paper we present observations of a hot star located at the center of an extremely faint nebula. We show that the star appears {\it not\/} to be in a post-AGB evolutionary stage. We conclude that the faint nebula is due simply to the star happening to pass through, and ionising, a region of the interstellar medium (ISM) having a relatively high density.

\section{The Ultra-Faint Nebula PN?\,G126.8$-$15.5 = F\lowercase{r} 2-30} \label{sec:ultrafaint}


\citet{Yuan2013} carried out a search for extremely faint PNe by examining some 1,700,000 target- and sky-fibre spectra from the Sloan Digital Sky Survey (SDSS), and seeking detections of the [\oiii] 4959--5007~\AA\ emission lines. For one of their nearly four dozen new candidate PNe, the [\oiii] lines were detected in the spectra of eight faint field stars lying across an area in Andromeda with a diameter of $\sim$0.35~degree, at a Galactic latitude of $-15\fdg5$. \citet{Yuan2013} designated the object PN?\,\,G126.8$-$15.5. The interrogation point indicates their uncertainty as to the nature of the extended object. They considered it to be a probable new PN, but could not rule out the possibility of a supernova remnant (SNR), based on the presence of relatively strong emission lines of [\sii] 6716--6731~\AA\ in some of the SDSS spectra. Their Figure~9 reproduces imagery from the Virginia Tech Spectral-Line H$\alpha$ Survey \citep{Dennison1998}, the {\it Wide-field Infrared Survey Explorer\/} \citep[\WISE;][]{Wright2010} at 22~\micron, and a broad-band blue photograph from the Space Telescope Science Institute Digitized Sky Survey.\footnote{\url{https://archive.stsci.edu/cgi-bin/dss_form}} All of these images show a faint, extended nebula at the site, with a major axis of at least $21'$.

\citet{Yuan2013} measured [\oiii] 5007~\AA\ surface brightnesses\footnote{The 5007\,\AA\ apparent magnitude of a source is defined  \citep{Jacoby1989} as $m_{5007} = -2.5 \log F_{5007} - 13.74$, where the emission-line flux $F_{5007}$ is in cgs units. It is approximately the apparent $V$ magnitude the source would have if the 5007~\AA\ line flux were distributed across the $V$ bandpass.} at the eight SDSS spectrum sites within the nebula, ranging from 26.9 to 28.9 mag\,arcsec$^{-2}$. The mean radial velocity (RV) they measured from the [\oiii] lines is $-25.3\pm2.6\,\kms$, with an rms scatter of $7.2\,\kms$ and a range of $-36.4$ to $-15.4\,\kms$.

This nebula is listed as a ``possible PN'' in the online Hong-Kong/AAO/Strasbourg/H$\alpha$ Planetary Nebulae (HASH) database\footnote{\url{http://hashpn.space/}} \citep{Parker2016}, but with a slightly different designation of PN G126.7$-$15.4. In the HASH catalogue, the object is assigned the name Fr~2-30, based on an  unpublished (to our knowledge) independent discovery of it by D.~Frew. For convenience, we will use this designation hereafter.

\section{Deep Imaging of F\lowercase{r} 2-30} \label{sec:deepimaging}

Long exposures on Fr 2-30 were obtained by J.T. with a Stellarvue SVX 152T 6-inch refractor\footnote{\url{https://www.stellarvue.com/svx152t/}} at his home observatory\footnote{For details see \url{http://starscapeimaging.com/}} in Ocean Springs, Mississippi, USA\null. A ZWO ASI 6200 CMOS camera\footnote{\url{https://astronomy-imaging-camera.com/product/asi6200mm-pro-mono}} at the $f$/8 focus provided an image scale, using $2\times2$ binning, of $1\farcs28$\,pixel$^{-1}$. The field of view (FOV) of the camera is $1\fdg70\times1\fdg13$. Narrow-band exposures were obtained in Astrodon\footnote{\url{https://farpointastro.com/collections/astrodon}} H$\alpha$ and Chroma\footnote{\url{https://www.chroma.com/}} [\oiii] 5007\,\AA\ filters, accumulating 42 1200~s exposures in each bandpass, totaling 14~hours each. These were supplemented with images in broadband Chroma Red, Green, and Blue filters, consisting of 20 300~s exposures in each. The grand total exposure time was 33~hours. These observations were obtained between 2021 September~10 and October~26. A final image was rendered by using {\tt PixInsight}\footnote{\url{https://pixinsight.com/}} to stack all of the frames. H$\alpha$ was mapped to red and [\oiii] to green and blue, while the broadband frames (showing field stars) were assigned to red, green, and blue.



\begin{figure*}
\centering
\includegraphics[width=0.85\textwidth]{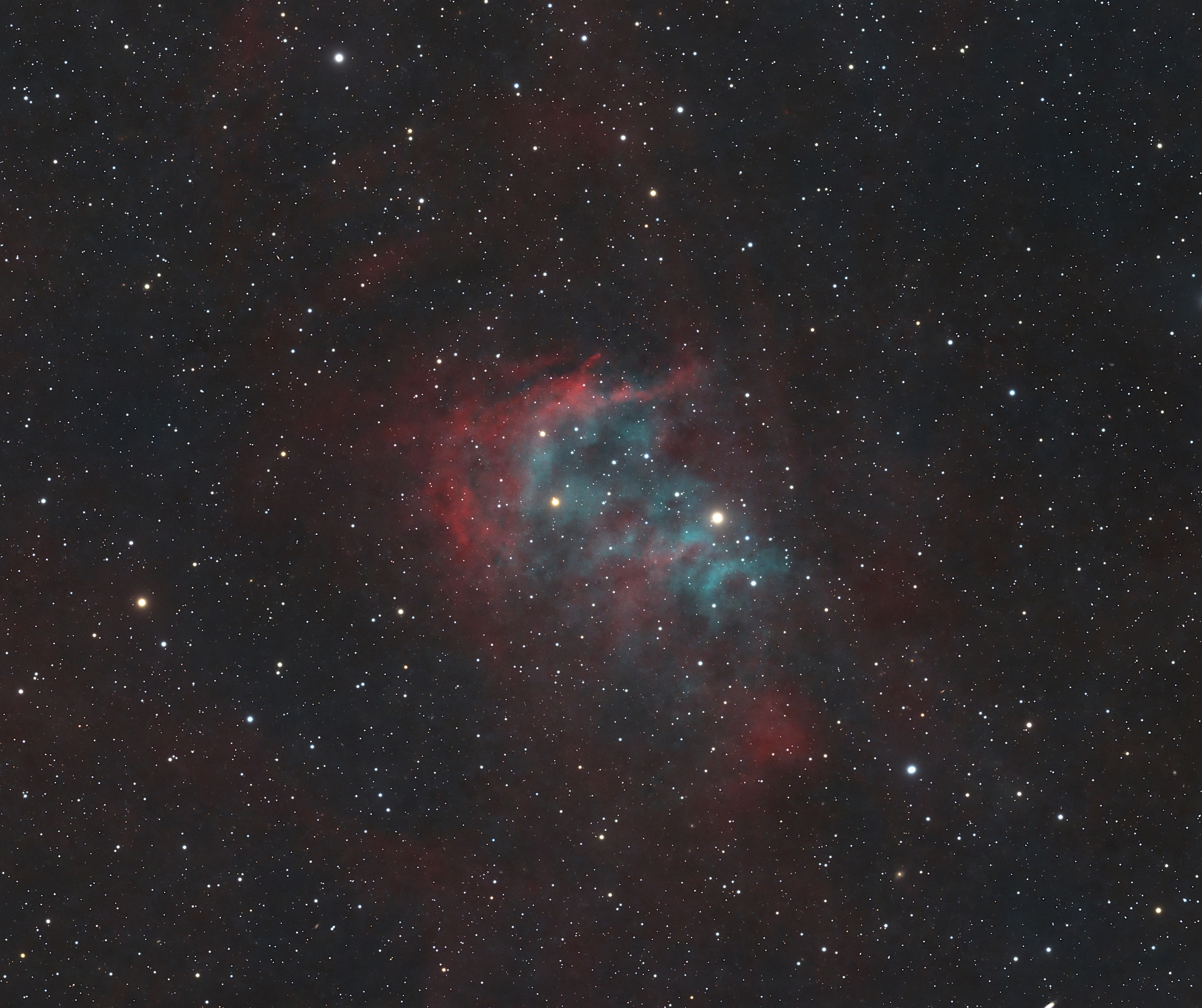}
\caption{Image of Fr 2-30 from Stark Bayou Observatory. \Ha+[\nii] is mapped to red, and [\oiii] 5007\,\AA\ to green and blue. Broadband frames of the stellar field are mapped to red, green, and blue. Height of frame $45'$. North on top, east on left. The ionising central star is identified below in Figure~\ref{fig:Fr2-30chart}.
\label{fig:Fr2-30image} }
\end{figure*}

Figure~\ref{fig:Fr2-30image} presents the resulting deep image. The nebula is seen to have an irregular and patchy, approximately elliptical morphology, elongated roughly from northeast to southwest. The dimensions of the brighter regions are about $22'\times14'$ in \Ha\ and about $14'\times9'$ in [\oiii], but these are extremely rough estimates because the edges are not well defined. Emission in the \Ha\ bandpass is especially prominent on the northeast edge of Fr~2-30. However, it should be noted that the Astrodon filter's bandpass is $\sim$50~\AA\ wide, and thus it also has sensitivity to the nebular [\nii] emission lines at 6548 and 6583~\AA\null. Inspection of the SDSS spectra of the nebula \citep{Yuan2013} shows that there is a significant, and sometimes dominant, contribution from [\nii] at these locations. 


\begin{figure*}
\centering
\includegraphics[width=0.85\textwidth]{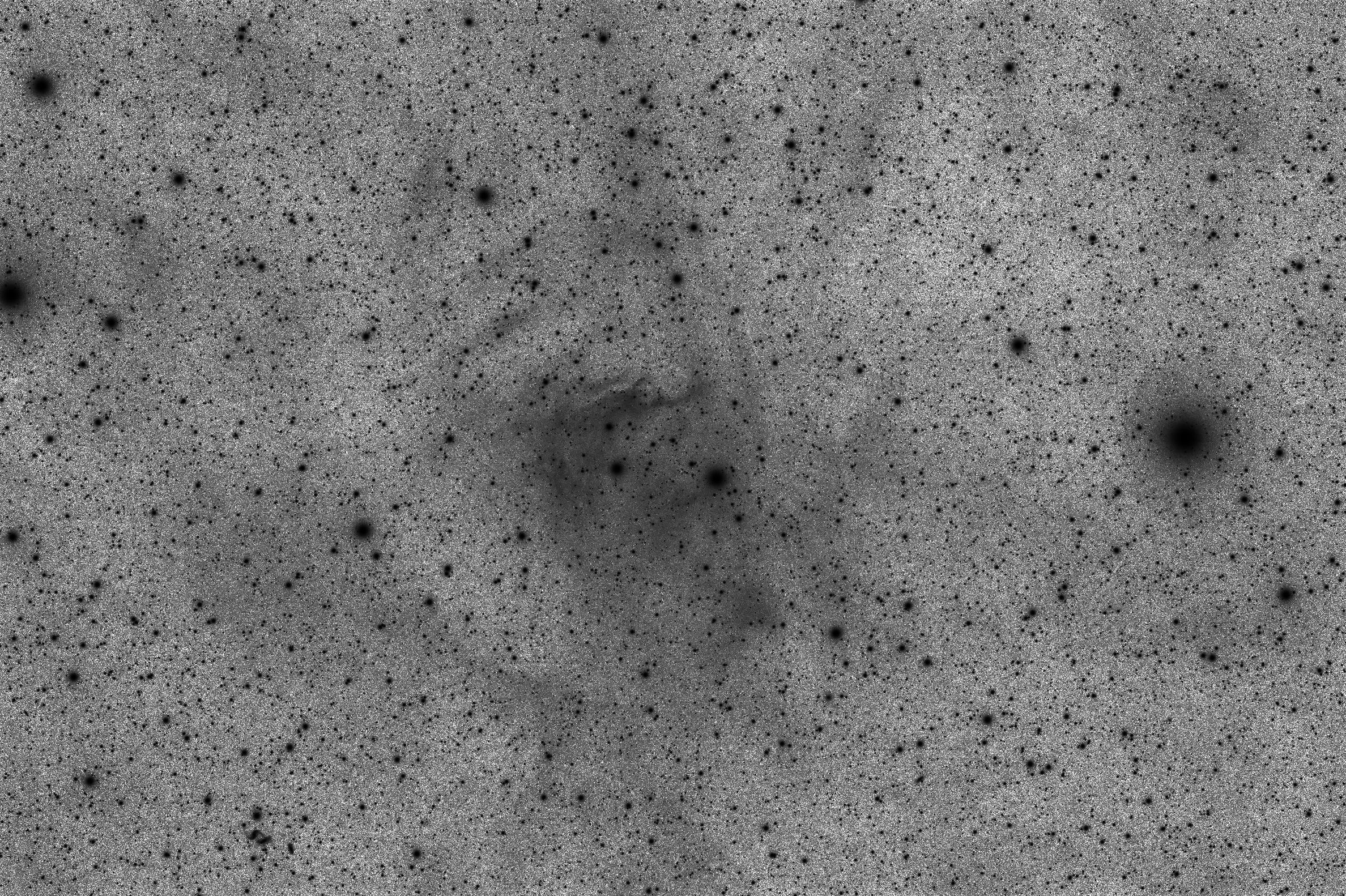}
\caption{Highly stretched image in the \Ha\ bandpass of the region surrounding Fr 2-30, showing the entire $1\fdg70\times1\fdg13$ camera field.  
\label{fig:DeepImage} }
\end{figure*}

Fainter \Ha\ emission is seen over much of the field shown in Figure~\ref{fig:Fr2-30image}, especially to the northeast, east, and southeast. To illustrate this more clearly, Figure~\ref{fig:DeepImage} shows a negative image of the entire $1\fdg70\times1\fdg13$ field in the \Ha\ filter, heavily stretched to show the faintest features. Now we see that a network of faint, patchy emission covers the entire field, and it appears that the Fr~2-30 nebula is a region of enhanced surface brightness within this larger extended structure. A deep, wide-angle, but low-spatial-resolution, \Ha\ image\footnote{Available at \url{http://www1.phys.vt.edu/~halpha/}} from the Virginia Tech Survey mentioned in \S\ref{sec:ultrafaint} shows that faint emission continues to extend over at least several degrees to the north-northwest and south-southeast.

\section{The Central Star} \label{sec:centralstar}

\begin{figure}
\centering
\includegraphics[width=0.45\textwidth]{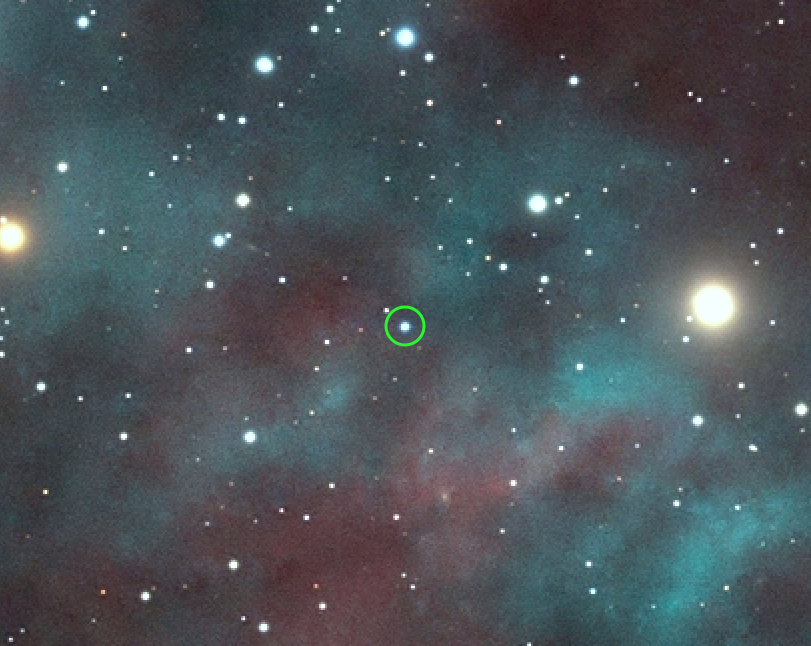}
\caption{Inset from Figure~\ref{fig:Fr2-30image}, showing the central star of Fr 2-30, encircled in green. Height of frame $410''$. North on top, east on left. 
\label{fig:Fr2-30chart} }
\end{figure}

Inspection of the recent \Gaia\/ Data Release~3 catalogue\footnote{\url{https://vizier.cds.unistra.fr/viz-bin/VizieR-3?-source=I/355/gaiadr3}} (DR3; \citealt{Gaia2016, Gaia2022}) showed that a 14th-magnitude blue star lies near the center of Fr~2-30. Figure~\ref{fig:Fr2-30chart} presents an inset from Figure~\ref{fig:Fr2-30image}, in which this star is marked. Properties of the star from \Gaia\/ DR3 are given in Table~\ref{tab:DR3data}, including its celestial and Galactic coordinates, parallax and proper motion, and the \Gaia\/ $G$ magnitude and $G_{\rm BP}-G_{\rm RP}$ colour. Given the star's extremely blue colour and its location very close to the center of the [\oiii] emission in Fr~2-30, there can be little doubt that it is responsible for excitation of the nebula. Moreover, the presence of this star, as well as the nebular morphology, appear to rule out the nebula being a SNR\null. The \Gaia\/ parallax gives a distance of the star (and of the nebula) of $890\pm25$~pc. The star's interstellar extinction, determined using the online tool\footnote{\url{https://stilism.obspm.fr/}} of \citet{Capitanio2017}, is $E(B-V)=0.094$. The resulting absolute $G$ magnitude is +4.3, confirming that it is a subluminous hot star. Its tangential space velocity is $66\,\kms$ at a position angle of approximately $273^\circ$ (i.e., almost exactly to the right in Figures~\ref{fig:Fr2-30image}, \ref{fig:DeepImage}, and~\ref{fig:Fr2-30chart}).  The $22'\times14'$ dimensions of the brightest parts of the nebula correspond to a physical size of about $5.7\times3.6$~pc.

\begin{table}
\centering
\caption{\Gaia\/ DR3 Data for Central Star of Fr 2-30.}
\label{tab:DR3data}
\begin{tabular}{lc} 
	\hline
Parameter & Value \\
	\hline
RA (J2000) & 01 13 16.941 \\
Dec (J2000) & +47 11 29.82 \\
$l$ [deg] &  126.78 \\
$b$  [deg] &  $-15.52$ \\
Parallax [mas] & $1.123\pm0.033$ \\
$\mu_\alpha$ [mas\,yr$^{-1}$] & $-15.531\pm0.036$ \\
$\mu_\delta$ [mas\,yr$^{-1}$] & $+0.876\pm0.022$ \\
$G$ [mag] &  14.35 \\
$G_{\rm BP}-G_{\rm RP}$ [mag] & $-0.37$ \\
 	\hline
\end{tabular}
\end{table}

A literature search revealed that this star had already been noted in a survey of the \Gaia\/ DR2 database by \citet{Geier2019} and \citet{2020A&A...635A.193G}, aimed at finding and cataloguing hot subdwarfs. These are defined as hot stars with absolute magnitudes placing them below the main sequence but brighter than the white-dwarf (WD) sequence in the HR diagram; for an exhaustive review of hot subdwarf stars, see \citet{Heber2016}. The \citet{Geier2019} and \citet{2020A&A...635A.193G} catalogues contain the Fr 2-30 central star, with \Gaia\/ ID 401413450281523584.	   

\citet{Lei2020} cross-matched the \citet{Geier2019} catalogue with data from the Large Sky Area Multi-Object Fibre Spectroscopic Telescope (LAMOST) survey, which has obtained moderate-resolution (resolving power $R=1800$) spectra for millions of stars. The wavelength coverage of the LAMOST spectra is 3800--9100~\AA\null. From this material, \citet{Lei2020} derived atmospheric parameters for 182 hot subdwarfs, including the central star of Fr~2-30. Their analysis resulted in a spectral type of subdwarf~O (sdO), stellar parameters of $\Teff=48\,680\pm820$~K and $\log g=6.06\pm0.04$, and a low helium content of $\log (n_{\rm He}/n_{\rm H})=-3.21\pm0.33$.

None of the works cited in this section mentioned the apparent association of the star with the Fr 2-30 nebula, which at present is not listed in the widely used SIMBAD\footnote{\url{http://simbad.u-strasbg.fr/simbad/}} database.

\section{HET LRS2 Observations} \label{sec:lrs2obs}

Prompted by the discovery of the faint nebula by \citet{Yuan2013} (see \S\ref{sec:ultrafaint}), and our identification of the central star (\S\ref{sec:centralstar}), we added the star to the target list for the HET/LRS2 spectroscopic survey described in \S\ref{sec:intro} and in more detail in Paper~I.

LRS2 provides integral-field-unit (IFU) spectroscopy with 280 $0\farcs6$-diameter lenslets that cover a $12''\times6''$ FOV on the sky. LRS2 is composed of two arms: blue (LRS2-B) and red (LRS2-R). All of our observations are made with the LRS2-B IFU\null. The LRS2-B arm employs a dichroic beamsplitter to send light simultaneously into two spectrograph units: the ``UV'' channel (covering 3640--4645~\AA\ at $R=1910$), and the ``Orange'' channel (covering 4635--6950~\AA\ at $R=1140$). Our LRS2-B observations of the Fr~2-30 central star were obtained on 2022 July~17 (exposure time 248~s) and 2022 November~18 (407~s).

The raw LRS2 data were initially processed with \texttt{Panacea}\footnote{\url{https://github.com/grzeimann/Panacea}}, which carries out bias and flat-field correction, fibre extraction, and wavelength calibration. Initial absolute flux calibration comes from default response curves and measures of the mirror illumination as well as the the exposure throughput from guider images.  We then used \texttt{LRS2Multi}\footnote{\url{https://github.com/grzeimann/LRS2Multi}} to perform sky subtraction, source modeling, and {1-D} spectral extraction.  We identified the locations of the central star in the two separate exposures, and subtracted from the stellar spectrum the median spectrum of all ``sky'' fibres more than $3\farcs5$ away.   The target star was modeled with a \citet{Moffat1969} spatial profile with $\beta = 3.5$, and we applied a weighted optimal extraction \citep{Horne1986} for each channel.  We resampled each channel's wavelength solution to a common linear grid with a 0.7~\AA\ spacing. Before combining the two nights' data, we applied a normalisation to account for an absolute-calibration offset. The adopted normalisation factors were 1.15 and 0.85 for 2022 July~17 and November~18, respectively.  We then combined the two nights' spectra, using a signal-to-noise (S/N)-weighted average. Finally we rectified the spectrum to a flat continuum for detailed analysis, as described below. 


The median sky spectrum that we derived is the sum of emission from the night sky and from the diffuse nebulosity surrounding the star within the IFU FOV\null. We extracted the nebular 
[\oiii] 5007~\AA\ emission line from the sky spectrum as follows. Because of moonlight, the night-sky brightnesses for 2022 July~17 and November~18 differed by a factor of ten, the latter being the fainter.  We used \texttt{PPXF} \citep{Cappellari2022} to model the night-sky continuum from 4910 to 5110~\AA, and subtracted the model in order to isolate the [\oiii] nebular line.
Despite the order-of-magnitude difference in night-sky continuum brightness, we found consistent measurements of the [\oiii] 5007~\AA\ emission from the two spectra. The S/N of the detections were 12 and 21 on the two nights, at a mean 5007~\AA\ surface brightness of $27.06\pm0.05$ mag arcsec$^{-2}$, as defined in \S\ref{sec:ultrafaint}.  The heliocentric RVs of the [\oiii] lines were measured to be $-12.2\pm11.2$ and $-15.6\pm5.4\,\kms$, for a weighted mean RV of $-15.0\pm4.9\,\kms$. Both the [\oiii] 5007~\AA\ surface brightness and RV obtained from our HET spectra are reasonably consistent with the values given at several nearby locations in the extended nebula by \citet[][see \S\ref{sec:ultrafaint}]{Yuan2013}, based on SDSS spectra.

\section{Atmospheric Analysis} \label{sec:atmospheric_analysis}

We used the T\"ubingen Model-Atmosphere Package \citep[TMAP;][]{Werner2003} to build a grid of non-LTE plane-parallel model atmospheres in radiative and hydrostatic equilibrium, containing hydrogen and helium. We used the model spectra to determine the effective temperature, surface gravity, and He/H abundance ratio of the Fr~2-30 central star. The grid spacings for \Teff\ and \logg\ were 5000\,K and 0.2\,dex, respectively. The He/H number ratio ranged from zero to 0.01 in initial steps of 0.001, with a finer spacing being used close to the finally determined value. For the best-fitting model, chosen by eye, we find \Teff $= 60\,000\pm5000$\,K, $\log g = 6.0\pm0.2$, and $n_{\rm He}/n_{\rm H} = 0.0017\pm0.0010$.

\begin{figure}
\begin{center}
\includegraphics[width=\linewidth]{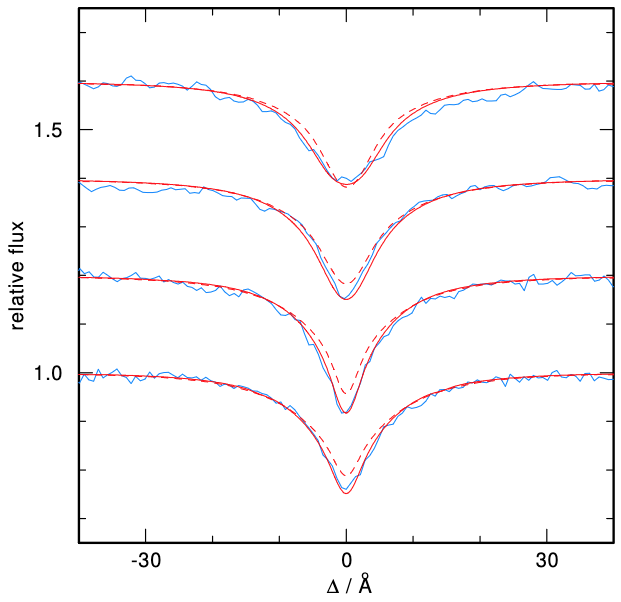}
\caption{
Effect of metal-line blanketing on the Balmer lines H$\alpha$ to H$\delta$ (from top to bottom). Dashed red lines are theoretical profiles from a pure hydrogen and helium model, while the solid red lines are from a model also containing metals, as explained in the text. The blue line is our observed spectrum of Fr~2-30.
  }
\label{fig:balmer}
\end{center}
\end{figure}

Close inspection of the best-fitting model spectrum from our grid of hydrogen-plus-helium models showed that the cores of the computed H$\beta$ -- H$\delta$ lines did not quite match the observation; the cores in the model spectrum were not deep enough. We considered this likely to be due to neglect in the model of metal-line blanketing, which can cause cooling of the outer atmospheric layers and potentially deepen the Balmer line cores \citep[e.g.,][]{1996ApJ...457L..39W}. Optical spectra of sdOs often look ``clean,'' in the sense that we see only hydrogen and helium lines. The problem is that because of the high effective temperatures, the metals are highly ionised and can usually be detected only in the ultraviolet. While helium is often depleted because of gravitational settling (as in the case of the Fr~2-30 sdO), radiative levitation can cause strong overabundances (compared to solar) of heavy elements, in particular iron and nickel. For example, ultraviolet observations by \citet{2018A&A...609A..89L} indicated extreme overabundances of Fe and Ni in four sdO stars that have very similar temperature and gravity to the sdO central star of Fr~2-30. We therefore computed a model having the Fr~2-30 temperature, gravity, and helium abundance, but this time including metals with abundances measured in the sdO star AGK$+81^\circ266$, whose parameters are quite similar, namely $T_{\rm eff} = 61\,860$\,K, $\log g = 6.07$, and $n_{\rm He}/n_{\rm H} = 0.0013$. We included the CNO elements, silicon, phosphorus, sulfur, iron, and nickel with the following abundances, given as number ratio relative to hydrogen: C = N = O = $1\times10^{-6}$, Si = $1\times10^{-5}$, P = $3\times10^{-6}$, S = $5\times10^{-6}$, Fe = $8\times10^{-4}$, Ni = $1\times10^{-4}$. 

Figure~\ref{fig:balmer} compares the observed Balmer-line profiles with those from model spectra with and without metals, for \Ha\ through H$\delta$. As expected, the metal-line-blanketed model provides a much better fit to the observed line cores than does the metal-free model. 

\begin{figure*}
 \centering  \includegraphics[width=0.9\textwidth]{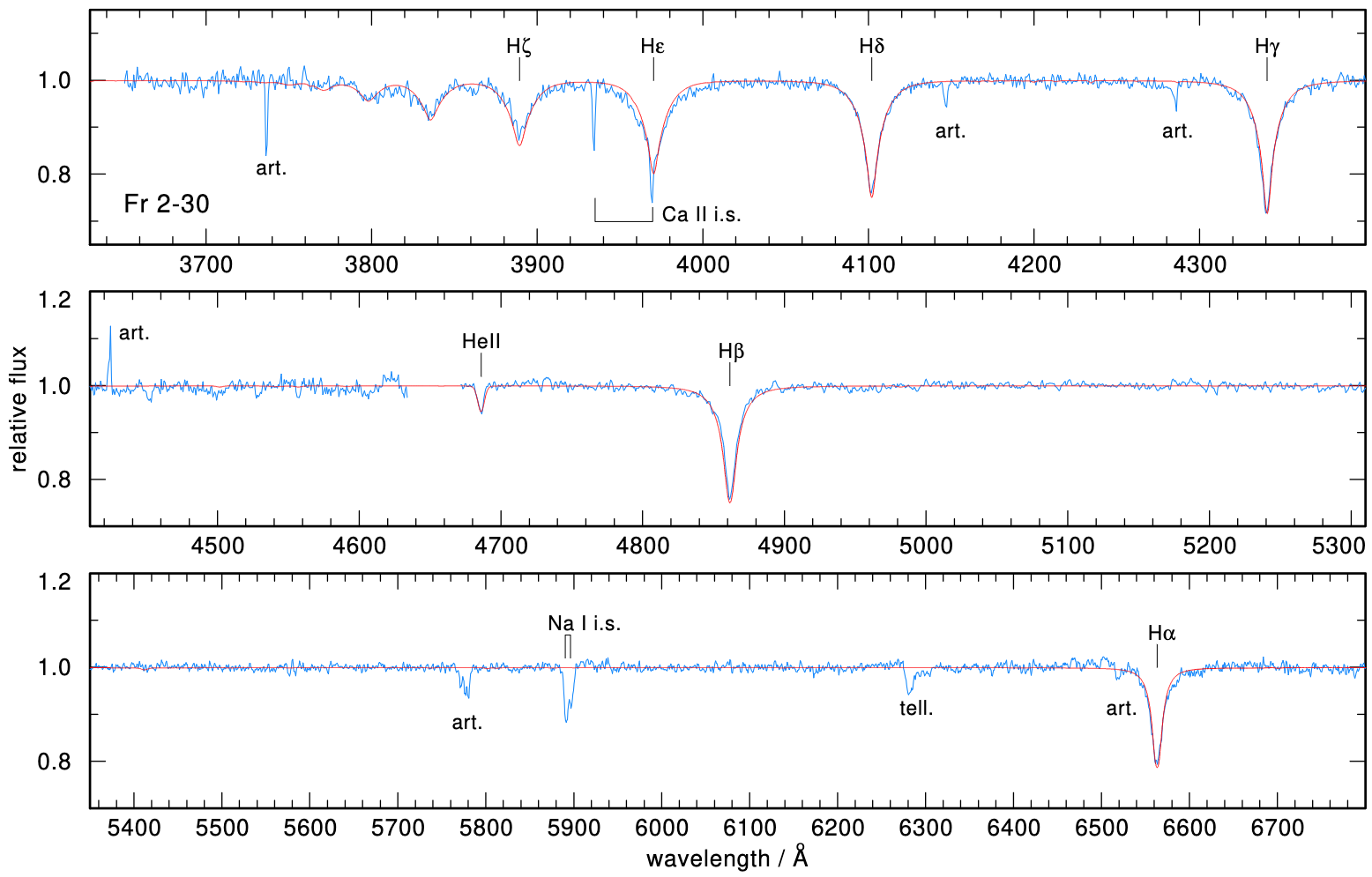}
 \caption{Observed spectrum of the Fr~2-30 central star (blue line), superposed with our final theoretical model (red line) with \Teff = 60\,000\,K, \logg = 6.0, $n_{\rm He}/n_{\rm H} = 0.0017$ (number ratio). The model includes metals (C, N, O, Si, P, S, Fe, Ni) with abundances as given in the text---but they do not produce any detectable absorption lines in the optical region. Identified photospheric lines are marked. Other labels are ``art.'' indicating artefacts, ``i.s.'' interstellar lines, and ``tell.'' telluric features. 
 }
\label{fig:spectrum}
\end{figure*}

In Figure~\ref{fig:spectrum} we plot the synthetic spectrum computed for the adopted stellar parameters, and including the metal abundances as listed above, as a red curve. It shows the excellent agreement with the normalised HET/LRS2-B spectrum, plotted as a blue curve. As noted above, no metallic lines are visible in the optical spectrum, in spite of their presence in the model and presumably in the star. Even for helium, only one line is detected, \heii\ 4686~\AA, in accordance with the very low helium abundance. 

Although our derived surface gravity agrees with the LAMOST result of \citet{Lei2020} cited above (\S\ref{sec:centralstar}), our effective temperature is higher by more than 10\,000\,K\null. One reason could be that the non-LTE models used by \citet{Lei2020} do not include metals. To further investigate this discrepancy we downloaded\footnote{From \url{http://www.lamost.org/dr8/v2.0/search}} the two available LAMOST spectra; one has a much higher S/N than the other one, and we inspected it in more detail.
There appears to be a problem with the flux calibration in the vicinity of the overlapping high-order Balmer lines, because the continuum flux is not smooth there. Thus the determination of the continuum level is uncertain. Since the Balmer decrement is temperature sensitive, a systematic error in the temperature determination may have resulted. We then compared the lower-order Balmer lines in the LAMOST spectrum with two models, one with our temperature of 60\,000\,K, and the other with 49\,000~K (close to the LAMOST value). The wings of the H$\gamma$ through H$\epsilon$ lines of the cooler model are too broad and fit worse than does the hotter model. A similar effect is seen at H$\beta$, but to a lesser extent. Comparison of the two models with our HET spectrum shows that the cooler model predicts slightly too strong profiles for the high-order Balmer lines. In the end we can only speculate that the automatic line fitting of the LAMOST analysis by \citet{Lei2020} could have been affected by flux-calibration issues. We also note that the small uncertainties given in the LAMOST analysis are only formal statistical errors, and do not include systematic errors. 

\section{Evolutionary Status} \label{sec:evolutionarystatus}

As described in \S\ref{sec:intro}, central stars of PNe are generally considered to be the cores of AGB stars that have ejected their outer layers and are evolving toward the onset of the WD cooling sequence. However, the atmospheric parameters of the exciting star of Fr~2-30---in particular its relatively low temperature and relatively high surface gravity---appear to be inconsistent with a post-AGB evolutionary status, as we explain below.  We therefore consider two alternative possibilities for its evolutionary state.

\begin{figure}
\begin{center}
\includegraphics[width=\linewidth]{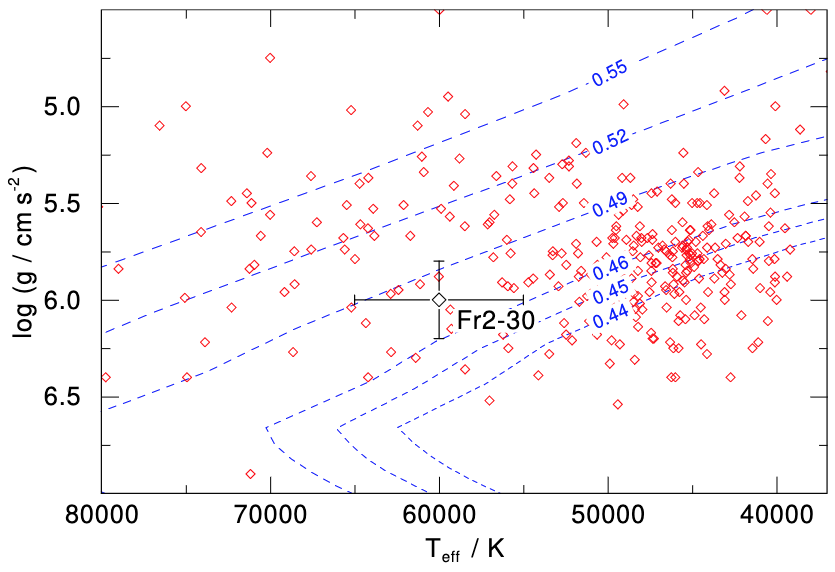}
\caption{Position of the exciting star of Fr~2-30 in the $\log g$ --
\Teff\ 
  diagram (black open circle with error bars) together with the field sdO stars in the \citet{2020A&A...635A.193G} catalogue with known atmospheric parameters (open red circles). Dashed lines are post-EHB evolutionary tracks, labeled with masses in solar units, from \citet{1993ApJ...419..596D}.
  }
\label{fig:gteff}
\end{center}
\end{figure}

Figure~\ref{fig:gteff} shows the position of the central star of Fr~2-30 in the ``Kiel diagram,'' $\log g$ versus \Teff. Also plotted are all of the known field sdO stars contained in the \citet{2020A&A...635A.193G} catalogue for which \Teff\ and $\log g$ values have been determined, and which lie within the ranges of the figure. (Note that the vast majority of these objects are {\it not\/} known to be central stars of PNe.) The figure shows that Fr~2-30 has parameters typical of the sdO population. Subdwarf~O stars are usually considered to be evolved low-mass helium-core burning stars lying at the blue end of the horizontal branch (HB) in the HR diagram, and defining the extreme HB (EHB). They also include post-EHB stars that have evolved away from the EHB as helium-shell burners. These objects are about to become carbon/oxygen-core WDs, thus avoiding the AGB stage entirely \citep[AGB-manqu\'e stars;][]{1990ApJ...364...35G}. Theoretical post-EHB evolutionary tracks connect the subdwarf B (sdB) stars on or near the EHB band to the helium-deficient sdO stars. In Figure~\ref{fig:gteff} we superpose a set of such tracks, computed by 
\citet{1993ApJ...419..596D}. In this interpretation of Fr~2-30 as a post-EHB star, we derive a mass of $0.48\pm0.02 \,M_\odot$.

EHB stars are thought to be the outcome of close-binary evolution. Strong mass loss of a star at the tip of the red-giant branch (RGB) strips most of the hydrogen-rich envelope via common-envelope (CE) evolution or Roche-lobe overflow, just before core helium ignition. Helium burning on the EHB lasts so long that a possible PN ejected at the end of the RGB phase has enough time to disperse into the ISM\null. Thus, if our star is in a post-EHB stage, it would be unlikely to host a detectable PN.

However, a second evolutionary scenario is conceivable. Fr~2-30 could be a star that left the RGB well {\it before\/} core helium ignition, because of strong mass loss during a CE phase. Such post-RGB stars are {\it hydrogen}-shell burners, which are about to become helium-core WDs. Thus they completely bypass the HB, evolving directly from the RGB towards the hot end of the WD cooling sequence. As in the case of the EHB stars, the ejected CE could briefly be visible as an ionised PN, until it dissipates.

\begin{figure}
\begin{center}
\includegraphics[width=\linewidth]{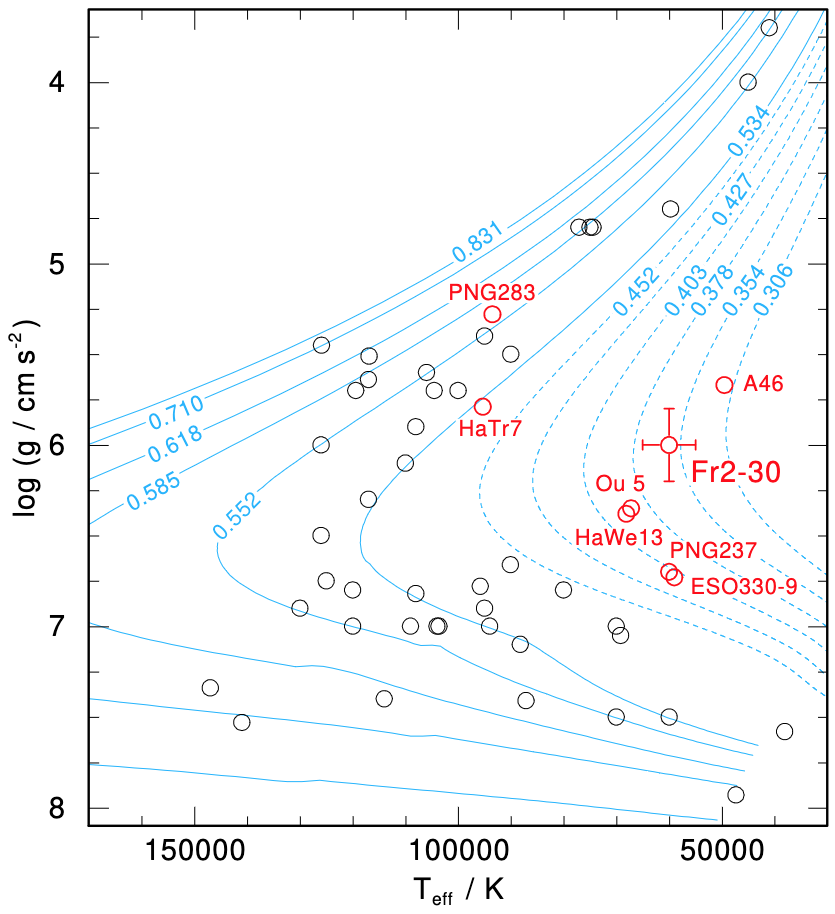}
\caption{
Position of the exciting star of Fr~2-30 in the $\log g$ --
\Teff\ diagram (red circle with error bars). H-rich post-AGB central stars of PNe are plotted as black circles. Their parameters are from
\citet{1999A&A...350..101N},
\citet{2007A&A...470..317R},
\citet{2010ApJ...720..581G},
\citet{2011MNRAS.417.2440H},
\citet{2012PhDT.......152Z},
\citet{2023MNRAS.519.2321J},
and Weidmann et al.\ (in press). Positions of PN central stars that are post-RGB  candidates are plotted as red circles; these are taken from \citet{Jones2022,Jones2023}, and Weidmann et al.\ (in press). Evolutionary tracks for post-AGB remnants (solid blue lines) by \citet{millerbertolami16} and for post-RGB remnants (dashed blue lines) by \citet{Hall2013} are labelled with the stellar masses in solar units.  
}
\label{fig:gteff_post-rgb}
\end{center}
\end{figure}

In Figure~\ref{fig:gteff_post-rgb} we again plot the position of the exciting star of Fr~2-30 in the $\log g$ -- \Teff\ diagram. This time we superpose (solid blue lines) a set of post-AGB evolutionary tracks computed by \citet{millerbertolami16}. Also plotted, as open black circles, are the parameters of a set of hydrogen-rich central stars of PNe, from the sources listed in the caption. Most of these central stars are seen to have atmospheric parameters consistent with post-AGB evolution. In contrast, Fr~2-30 lies at a position clearly in conflict with a post-AGB status. Also superposed (dashed blue lines) is a set of post-RGB tracks from 
\citet{Hall2013}. Fr~2-30 does lie in the region of these post-RGB tracks. Its position yields a mass of $0.36\pm0.02\,M_\odot$, lower than in the post-EHB case.

According to the tracks, the post-RGB age of Fr~2-30 is $t_{\rm evol}=400^{+80}_{-180}$\,kyr.  If we assume this age, a distance of 890~pc, and an expansion velocity of $20\kms$, the material ejected during the CE interaction would now have an angular diameter of roughly $1\fdg1$---which is in fact similar to the angular scale of the brighter portions of the nebulosity seen in Figures~\ref{fig:Fr2-30image} and~\ref{fig:DeepImage}. However, the nebulosity has the appearance of being part of a wider ambient network extending even further from the exciting star, and does not have the morphology typically seen in PNe.

Surprisingly, however, there {\it are\/} a few central stars of apparently genuine PNe that also occupy a location in the Kiel diagram similar to that of the Fr~2-30 exciting star. This is illustrated by the open red circles plotted in Figure~\ref{fig:gteff_post-rgb}, taken from recent papers by \citet{Jones2022,Jones2023} and Weidmann et al.\ (in press).  (Two of the plotted objects, PNG283 and HaTr\,7, are considered to be post-RGB stars because of low dynamical masses, rather than their locations in the figure.) These stars are known to be members of close binaries, making it likely that the surrounding nebulae are ejected CEs.

We can try to test the masses resulting from the two evolutionary scenarios by calculating the spectroscopic distance of the star, obtained from the relation
$$ d {\rm [pc]}= 7.11\times 10^{4} \sqrt{H_\nu\cdot M\cdot 10^{0.4V_0-\log g}}\ ,$$
where $H_\nu = 1.0\times10^{-3}\,\rm erg\,cm^{-2} \, s^{-1} \, Hz^{-1}$ is the Eddington flux of the model atmosphere at 5400\,\AA, $M$ is the stellar mass (in $M_\odot$), $V_0=V-A_V$ is the dereddened visual magnitude, $V$ is the observed magnitude, and $A_V$ is the visual extinction. Using the values $V=14.378$ \citep{2013AJ....145...44Z} and $A_V=3.1\times0.094$ (\S\ref{sec:centralstar}), we derive a distance of $d=1020^{+304}_{-236}$~pc for the case of a post-EHB star, or $d=883^{+263}_{-204}$~pc for the case of a post-RGB star. The errors are dominated by the uncertainty in the surface gravity. Both results are in  statistical agreement with the \Gaia\/ parallax distance of $890\pm25$~pc (see \S\ref{sec:centralstar}), so this test is not decisive.

\section{Is There a Binary Companion?}

For both the post-EHB and post-RGB evolutionary scenarios we would expect the Fr~2-30 central star to have had a close companion. The companion is needed to strip the envelope of the AGB-manqu\'e or RGB progenitor and place the remnant onto a post-EHB or post-RGB evolutionary track. Here we consider whether there is observational evidence for such a companion, { recognising also that it could have been disrupted or merged with the central star}.

We measured the RV of the central star using our two HET/LRS spectra. Centroid wavelengths of four Balmer lines, H$\alpha$ through H$\delta$, were measured and converted to velocities; the means were then corrected to heliocentric, all using standard tasks in {\tt IRAF}.\footnote{{\tt IRAF} was distributed by the National Optical Astronomy Observatories, operated by AURA, Inc., under cooperative agreement with the National Science Foundation.} Results are given in the final two rows of Table~\ref{tab:RVs}, where the errors are estimated from the scatter among the four hydrogen lines. 

We also determined RVs in the same manner for the two LAMOST spectra described in \S\ref{sec:atmospheric_analysis} (after adjusting from vacuum to air wavelengths). These results are listed in the first two rows of Table~\ref{tab:RVs}. The second LAMOST spectrum has a relatively low S/N, and the resulting RV is only approximate.\footnote{In a follow-up to the \citet{Lei2020} study discussed in our \S\ref{sec:centralstar}, \citet{Luo2020} used the LAMOST spectra to measure the RV of the star, finding $-107\pm31\,\kms$. This is apparently a mean of the RVs from the two observations; it agrees reasonably well with the average of our two independent measurements of the LAMOST spectra in Table~\ref{tab:RVs}, but with a larger stated uncertainty.}

The measurements in Table~\ref{tab:RVs} provide evidence that the RV is variable, but the uncertainties are especially large for two of the velocities. Additional measurements are needed to investigate the situation further.

\begin{table}
\centering
\caption{Heliocentric Radial Velocities of Fr 2-30 Central Star.}
\label{tab:RVs}
\begin{tabular}{lccl} 
	\hline
Date & HJD$-$2400000 & RV       & Spectrum \\
     &               & [$\kms$] & Source       \\
	\hline
2014 October 27  & 56958.178 & $-88.6\pm2.5$ & LAMOST \\
2017 November 20 & 58078.086 & $-110.1\pm16.2$ & LAMOST \\
2022 July 17     & 59777.908 & $-69.6\pm7.8$ & HET/LRS2 \\
2022 November 18 & 59901.809 & $-53.8\pm3.2$ & HET/LRS2 \\
  \hline
\end{tabular}
\end{table}

We also investigated whether the star is photometrically variable by downloading data from the All-Sky Automated Survey for Supernovae (ASAS-SN) website\footnote{\url{https://asas-sn.osu.edu}} \citep{Shappee2014, Kochanek2017} and from the {\it Transiting Exoplanet Survey Satellite\/} ({\it TESS}) database.\footnote{\url{https://archive.stsci.edu/missions-and-data/tess}} Neither data set showed evidence for optical variability in excess of the observational errors, on timescales of minutes to weeks.

Lastly we determined the spectral-energy distribution (SED) of the central star by obtaining photometric data from the following catalogues: optical data from the SDSS Data Release~12\footnote{\url{https://skyserver.sdss.org/dr12/en/home.aspx}} and the Panoramic Survey Telescope and Rapid Response System (Pan-STARRS) Data Release~2\footnote{\url{https://catalogs.mast.stsci.edu/panstarrs}}, and infrared data\footnote{\url{http://irsa.ipac.caltech.edu/frontpage}} from the Two-Micron All-Sky Survey (2MASS) and \WISE\null. (We note that these data are assembled conveniently in the \citealt{2020A&A...635A.193G} catalogue, which gives full literature references.)

\begin{figure}
\centering
\includegraphics[width=\linewidth]{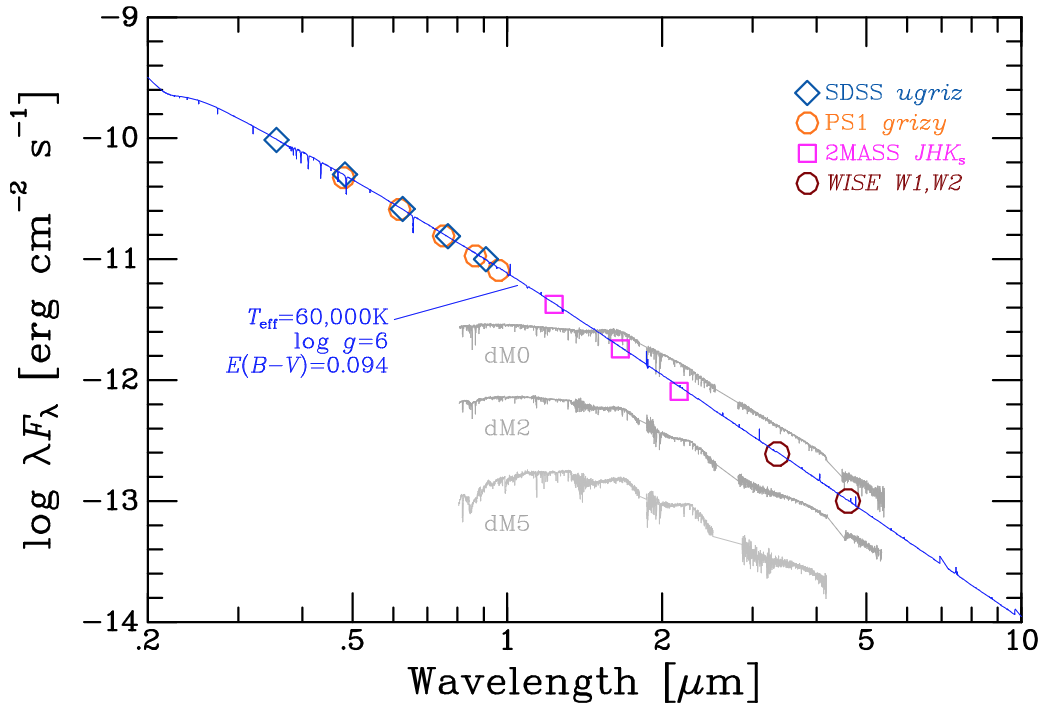}
\caption{Spectral-energy distribution of the Fr~2-30 central star from photometry in the SDSS, Pan-STARRS, 2MASS, and \WISE\/ surveys, colour-coded as indicated in the figure legend. Error bars are smaller than the plotting symbols. Superposed (blue curve) is a model-atmosphere spectrum calculated for the adopted parameters of the star, corrected for interstellar reddening. Also superposed (gray curves) are spectra of three nearby M dwarfs, scaled to the distance of Fr~2-30, as described in the text.  
\label{fig:sed} }
\end{figure}

The resulting SED for Fr~2-30 is plotted in Figure~\ref{fig:sed}. Superposed is the theoretical spectrum produced by TMAP for the adopted atmospheric parameters; it has been corrected for a reddening of $E(B-V)=0.094$, using the formulation of \citet{Cardelli1989}. The model-atmosphere spectrum agrees very well with the observations. There is no evidence at near-infrared wavelengths for the presence of a cool companion. To quantify this, we obtained calibrated spectra of three nearby M dwarfs from the NASA InfraRed Telescope Facility (IRTF) spectral library\footnote{\url{http://irtfweb.ifa.hawaii.edu/~spex/IRTF_Spectral_Library/}} \citep{Cushing2006,Rayner2009}. The three stars are HD~19305 (M0~V), HD~95735 (M2~V), and Gl~51 (M5~V), and their spectra, scaled to the distance of Fr~2-30, are plotted in Figure~\ref{fig:sed}. The comparison shows that a main-sequence companion earlier than about M2~V (with a mass greater than $\sim\!0.4\,M_\odot$) would have produced a detectable infrared excess at the \WISE\/ $W1$ and $W2$ bandpasses.\footnote{Unfortunately Fr\,2-30 was only marginally detected in the $W3$ and $W4$ bandpasses, at $\rm S/N\simeq1.5$ and 2.6, respectively.} Additionally, an even lower-mass binary companion lying sufficiently close to the hot primary for significant heating effects could be brightened enough to become detectable. But there is still a wide range of possible lower masses for a { main-sequence} companion, whose presence is suggested by the variable RV.
{ It is also possible that an unseen companion could be a WD, whose small radius would explain the apparent absence of irradiation effects.}

\section{True PN or Chance Encounter?}


\citet{FrewParker2010} have discussed PN ``mimics,'' which are emission nebulae that superficially imitate the appearance of a true PN, but are not actually material ejected from a low-mass hot central star. In particular, a  hot subdwarf that happens to pass through a sufficiently dense region of the ISM can locally create a Str\"omgren zone of ionised gas. \citet{FrewPHL9322010} describe the specific case of PHL~932, in which a sdB star is accompanied by an off-center ionised nebula. The authors present several lines of evidence that this is merely the result of a chance encounter of the hot star with high-Galactic-latitude molecular gas. They interpret the one-sided nebulosity as a ``wake'' of recombining plasma, and point out that the proper motion of the star is indeed in the direction away from the nebula. EGB~5 is another example of a sdB star lying at the edge of a faint emission nebula. In this case, the ionising star was found to be a 16.5-day single-lined spectroscopic binary by \citet{GeierEGB52011}. These authors discuss the possibility that the EGB~5 nebula is an ejected CE, but the off-center position of the star suggests a more likely explanation is a chance ISM encounter similar to PHL~932 (in which case the actual ejecta from the CE interaction dissipated long ago). { Yet another example is Fr~2-22, an extremely faint and little-studied nebula interpreted as a PN mimic by \citet{Hillwig2022}, who found the ionising star to be a sdB single-lined binary with an orbital period of only 0.255~day.} 

In the case of Fr~2-30, perhaps the strongest evidence against the nebula being ejecta from the central star is the discordance between the RVs of the nebula and the star. As noted in \S\ref{sec:lrs2obs}, we found a mean RV of the [\oiii] 5007~\AA\ emission line in the immediate vicinity of the star of $-15.0\pm4.9\,\kms$  from our two LRS2 spectra. The nebular RV was also measured at larger separations from the star in SDSS spectra by \citet{Yuan2013}, giving a mean of $-25.3\pm2.6\,\kms$ (see \S\ref{sec:ultrafaint} for details). These values are significantly different from the stellar RVs presented in Table~\ref{tab:RVs}, which range from $-53.8$ to $-110.1\,\kms$. 

In addition, a PN ejected long ago would be expected to have a hollow-shell morphology,\footnote{See, for example, a deep image of the ancient PN EGB~6 recently obtained by J.T.: \url{http://starscapeimaging.com/EGB6/EGB6.html}.} in contrast to the appearance of Fr~2-30 in Figure~\ref{fig:Fr2-30image}. In fact, the deep H$\alpha$ image shown in Figure~\ref{fig:DeepImage} strongly suggests that the nebula is simply the brightest part of an extensive network of faint material that covers most of the figure. 

Thus Fr~2-30 is most likely a PN mimic, created by the chance encounter of its central star with a relatively dense region of the ISM\null. Note that Figure~\ref{fig:Fr2-30image} shows enhanced red emission on the east and east-northeast edge of the nebula. As discussed in \S\ref{sec:deepimaging}, SDSS spectra exhibit strong [\nii] and [\sii] emission in this region, indicating a low level of excitation; thus the red colour could be largely due to strong [\nii] emission, which is included in the H$\alpha$ filter bandpass. We suggest that we are seeing a recombination wake, similar to that in PHL~932 discussed above---and, consistent with this picture, we note that the proper motion of the Fr~2-30 star is indeed in the western direction (to the right in the figure), as described in \S\ref{sec:centralstar}.




Finally, however, our discussion leaves unexplained the existence of several PN central stars in Figure~\ref{fig:gteff_post-rgb} that, like Fr~2-30, appear to lie on post-RGB evolutionary tracks. They are all known or suspected members of post-CE binaries \citep{Jones2022,Jones2023}. As in the case of Fr~2-30, the hot stars have evolutionary timescales that greatly exceed the dissipation timescale for a typical PN, but several of them at least are nevertheless surrounded by apparently genuine PNe. These objects remain an evolutionary puzzle, whether or not Fr~2-30 itself is merely a PN mimic.

\section{Summary}

We have investigated the nature of the extremely faint emission nebula PN?\,G126.8$-$15.5 = Fr~2-30 and its 14th-magnitude central star. We present a deep narrow-band image of the nebula, based on long exposures at \Ha\ and [\oiii] 5007~\AA\null. The image shows a diffuse nebula, fading into a surrounding extended network of faint emission. We obtained optical spectrograms of the central star, showing it to be a hot subdwarf~O star, and confirming that it is the source of excitation of the central portions of the nebula. Based on its \Gaia\/ parallax, the distance of the star (and the nebula) is 890~pc. A model-atmosphere analysis of the central star yields an effective temperature of $T_{\rm eff}=60,000$~K and a surface gravity of $\log g= 6.0$. The helium content is very low, $n_{\rm He}/n_{\rm H} = 0.0017$, a consequence of gravitational settling. However, we show that it is necessary to assume high abundances of heavy elements, up to iron and nickel, due to radiative levitation. This is needed in order to explain the observed line profiles of the hydrogen Balmer lines---even though no metal lines are actually visible in the optical spectrum. 

The position of the Fr~2-30 central star in the $\log g$ -- $T_{\rm eff}$ plane is inconsistent with the post-AGB evolutionary status typical of most PN nuclei. We consider instead two alternative evolutionary scenarios resulting from binary-star interactions: helium-burning post-extreme-horizontal-branch, or hydrogen-burning post-red-giant-branch. In either case the evolutionary timescales are so long that detectable ionised ejecta should not be present. 

We tentatively find evidence for a variable RV, suggesting that the Fr~2-30 central star may be a close binary. However, there is no evidence for photometric variations, and the SED rules out a main-sequence companion earlier than about spectral type M2~V\null. 

Finally we show that the RVs of the star and surrounding nebula are discordant. This, along with the morphology of the nebula, strongly suggests that Fr~2-30 is the result of a chance encounter between the hot sdO star and an interstellar cloud---a PN mimic. There appears to be a wake of recombining plasma on the east and east-northeast edge of the nebula, diametrically opposed to the direction of motion of the star determined from the \Gaia\/ proper motion. Left unexplained, however, is the apparent existence of several central stars of genuine planetary nebulae that have evolutionary states and long evolutionary ages similar to those of the Fr~2-30 nucleus.

\section*{Acknowledgements}


K.W. thanks Nicole Reindl for useful discussions. 

We thank the HET queue schedulers and nighttime observers at McDonald Observatory for obtaining the data discussed here.

The Hobby-Eberly Telescope (HET) is a joint project of the University of Texas at Austin, the Pennsylvania State University, Ludwig-Maximilians-Universit\"at M\"unchen, and Georg-August-Universit\"at G\"ottingen. The HET is named in honor of its principal benefactors, William P. Hobby and Robert E. Eberly.

The Low-Resolution Spectrograph 2 (LRS2) was developed and funded by the University of Texas at Austin McDonald Observatory and Department of Astronomy, and by Pennsylvania State University. We thank the Leibniz-Institut f\"ur Astrophysik Potsdam (AIP) and the Institut f\"ur Astrophysik G\"ottingen (IAG) for their contributions to the construction of the integral-field units.

We acknowledge the Texas Advanced Computing Center (TACC) at The University of Texas at Austin for providing high-performance computing, visualisation, and storage resources that have contributed to the results reported within this paper.

This work has made use of data from the European Space Agency (ESA) mission
{\it Gaia\/} (\url{https://www.cosmos.esa.int/gaia}), processed by the {\it Gaia\/}
Data Processing and Analysis Consortium (DPAC,
\url{https://www.cosmos.esa.int/web/gaia/dpac/consortium}). Funding for the DPAC
has been provided by national institutions, in particular the institutions
participating in the {\it Gaia\/} Multilateral Agreement.

This research has made use of the SIMBAD database, operated at CDS, Strasbourg, France.

The Large Sky Area Multi-Object Fiber Spectroscopic Telescope (LAMOST) is a National Major Scientific Project built by the Chinese Academy of Sciences. Funding for the project has been provided by the National Development and Reform Commission. LAMOST is operated and managed by the National Astronomical Observatories, Chinese Academy of Sciences. 

    Funding for the SDSS and SDSS-II has been provided by the Alfred P. Sloan Foundation, the Participating Institutions, the National Science Foundation, the U.S. Department of Energy, the National Aeronautics and Space Administration, the Japanese Monbukagakusho, the Max Planck Society, and the Higher Education Funding Council for England. The SDSS Web Site is http://www.sdss.org/.

    The SDSS is managed by the Astrophysical Research Consortium for the Participating Institutions. The Participating Institutions are the American Museum of Natural History, Astrophysical Institute Potsdam, University of Basel, University of Cambridge, Case Western Reserve University, University of Chicago, Drexel University, Fermilab, the Institute for Advanced Study, the Japan Participation Group, Johns Hopkins University, the Joint Institute for Nuclear Astrophysics, the Kavli Institute for Particle Astrophysics and Cosmology, the Korean Scientist Group, the Chinese Academy of Sciences (LAMOST), Los Alamos National Laboratory, the Max-Planck-Institute for Astronomy (MPIA), the Max-Planck-Institute for Astrophysics (MPA), New Mexico State University, Ohio State University, University of Pittsburgh, University of Portsmouth, Princeton University, the United States Naval Observatory, and the University of Washington.

The Pan-STARRS1 Surveys (PS1) have been made possible through contributions of the Institute for Astronomy, the University of Hawaii, the Pan-STARRS Project Office, the Max-Planck Society and its participating institutes, the Max Planck Institute for Astronomy, Heidelberg and the Max Planck Institute for Extraterrestrial Physics, Garching, The Johns Hopkins University, Durham University, the University of Edinburgh, Queen's University Belfast, the Harvard-Smithsonian Center for Astrophysics, the Las Cumbres Observatory Global Telescope Network Incorporated, the National Central University of Taiwan, the Space Telescope Science Institute, the National Aeronautics and Space Administration under Grant No.\ NNX08AR22G issued through the Planetary Science Division of the NASA Science Mission Directorate, the National Science Foundation under Grant No.\ AST-1238877, the University of Maryland, and Eotvos Lorand University (ELTE).

This paper also used data collected with the \TESS\/ mission, obtained from the MAST data archive at the Space Telescope Science Institute (STScI). Funding for the \TESS\/ mission is provided by the NASA Explorer Program. STScI is operated by the Association of Universities for Research in Astronomy, Inc., under NASA contract NAS 5-26555.

This research has made use of the NASA/IPAC Infrared Science Archive, which is operated by the Jet Propulsion Laboratory, California Institute of Technology, under contract with NASA.

This publication makes use of data products from the Two Micron All Sky Survey, which is a joint project of the University of Massachusetts and the Infrared Processing and Analysis Center/California Institute of Technology, funded by NASA and the NSF.

It also makes use of data products from the {\it Wide-field Infrared Survey Explorer}, which is a joint project of the University of California, Los Angeles, and the Jet Propulsion Laboratory/California Institute of Technology, and NEOWISE, which is a project of the Jet Propulsion Laboratory/California Institute of Technology. \WISE\/ and NEOWISE are funded by NASA.

\section*{Data Availability}




The HET/LRS2 spectra are available upon reasonable request to H.E.B. or G.R.Z.




\bibliographystyle{mnras}
\bibliography{ovi_pnni_refs} 








\bsp	
\label{lastpage}
\end{document}